\documentstyle[11pt,iau185,twoside,epsf]{article}
\def\edcomment#1{\iffalse\marginpar{\raggedright\sl#1\/}\else\relax\fi}
\reversemarginpar

\begin{document}
\title{
          Automated classification of variable stars for ASAS data
      }
 \author{
               Laurent Eyer \& Cullen Blake
        }
\affil{
        Princeton University Observatory, Princeton, NJ 08544, USA
      }
\begin{abstract}
With the advent of surveys generating multi-epoch photometry and their
discoveries of large numbers of variable stars, the classification of
the obtained times series has to be automated. We have developed a
classification algorithm for the periodic variable stars using a
Bayesian classifier on a Fourier decomposition of the light
curve. This algorithm is applied to ASAS (All Sky Automated Survey,
Pojmanski 2000). In the case of ASAS, 85\% of variable objects are red
giants. A remarkable relation between their period and amplitude is
found for a large fraction of those stars.
\end{abstract}
\vspace{-0.5truecm}
\section{Introduction}
\vspace{-0.2truecm}
In its test-implementation, the ASAS project measured 50 fields
($2\times 3$ deg$^2$ each) in $I$-band with a 135~mm f/1.8 telephoto
lens during the years 1997-2000. Pojmanski detected about 3900
variables stars, he listed among them 380 periodic variable stars.
We propose an automated method which classifies a subsample from ASAS
stars in a two step procedure by: 1) finding a satisfactory Fourier
decomposition for the light curve, 2) applying Autoclass (Cheesemen
1996), a Bayesian classifier, on the parameters obtained for each
light curve. Several tests were done and the best classification was
obtained when Period, Amplitude, Skewness and Amplitude ratio (first
overtone/fundamental amplitudes of the Fourier decomposition) were
used as the input parameters.  The subsample is formed by 458 stars
which have a fair periodic behaviour, and some time series with
aliasing periods have been removed.
\vspace{-0.5truecm}
\section{Results}
\vspace{-0.2truecm}
For a fraction of red giant stars, a clear relation between period and
amplitude can be seen (cf. Fig.~1, left). This relation is also seen
in infrared photometry (van Loon, these proceedings).
The classes found are (see Fig.~1, right):
   small amplitude and sinusoidal curves ($\sim$100),
   Eclipsing binaries ($\sim$144),
   Cepheids ($\sim$48),
   SARV ($\sim$40),
   SR ($\sim$81),
   Mira ($\sim$45).
The RR Lyrae stars are too few (too faint) to form a group, so they
might be recovered as extreme objects in some classes.
Some classes are divided in subgroups. For instance, the eclipsing
binaries are classified in three subgroups, which correspond
approximately to EA, EB, and EW, but with some mixture. The
decomposition in Fourier series is not optimal for such a
separation. Principal components analysis will be applied to separate
the different types of eclipsing systems.  The subgroups of SRs will
be studied to see if they corresponds to real physical distinctions.
\vspace{-0.5truecm}
\section{Conclusion}
\vspace{-0.2truecm}
With the method we propose on the ASAS sample, we show that an
Automated Classification can be reached with a level of incorrect
classification of about 5\%. This rate has to be reduced when very
large datasets will be considered. There are, of course, irreducible
classification ambiguities from the light curve alone (e.g. RRc and
eclipsing binaries of EW type unless measured with very accurate
photometry), but multi-colour photometry and/or spectroscopy can help
to lift the ambiguity.

Our acknowledgements go to Prof. B.Paczynski, Dr C.Alard \& Dr
A.Gautschy for their fruitful discussions and comments.
\vspace{-0.5truecm}
\section{Internet Links}
\vspace{-0.2truecm}
\noindent
ASAS Home Page: http://archive.princeton.edu/$^\sim$asas/\\
For this work: http://www.astro.princeton.edu/$^\sim$leyer/ASAS/\\
See also HAT Home Page: http://www.astro.princeton.edu/$^\sim$bakos/HAT

%----------------------------- FIG.1  -------------------------------------
\begin{figure}
\plotfiddle{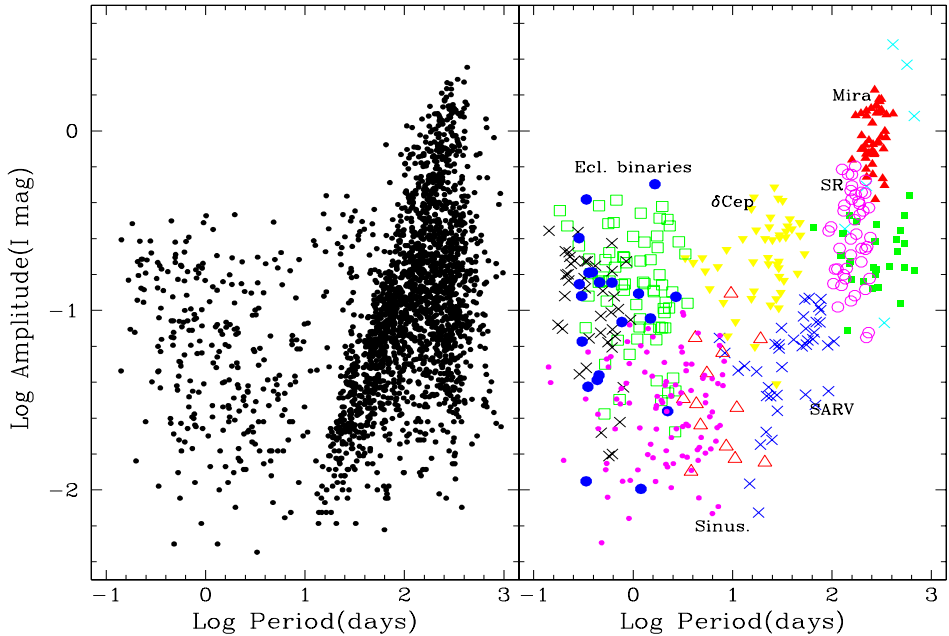}{7.8cm}{-90}{50}{45}{-200}{255}
\caption{\label{fig:colour}
                Diagram of amplitudes versus periods. Left: raw diagram
                (the periods are obtained with the Lomb algorithm, the
                whole sample is presented with the exception of some stars
                with aliasing periods).
                Right: Result obtained after the modeling of the Fourier
                decomposition. The main classes obtained are written
                next to the points.
        }
\end{figure}
%--------------------------------------------------------------------------
%

\end{document}